\documentclass[twocolumn,osajnl,preprintnumbers,showpacs]{revtex4}
\usepackage{graphicx}
\usepackage{amssymb,amsmath,textcomp,epsfig,subfigure}

\begin{document}
\date{\today}

\title{Multiple scattering suppression in dynamic light scattering based on a digital camera detection scheme}
\author{Pavel Zakharov}
\email{Pavel.Zakharov@unifr.ch}
\author{Suresh Bhat}
\author{Peter Schurtenberger}
\author{Frank Scheffold}
\affiliation{Physics Department, University of Fribourg, CH-1700
Fribourg, Switzerland}
\homepage{www.unifr.ch/physics/mm/}

\begin{abstract}

We introduce a charge coupled device (CCD) camera based detection scheme in dynamic light
scattering that provides information on the single-scattered
auto-correlation function even for fairly turbid samples. It is
based on the single focused laser beam geometry combined
with the selective cross correlation analysis of the scattered
light intensity. Using a CCD camera as a multispeckle detector we
show how spatial correlations in the intensity pattern can be
linked to single and multiple scattering processes. Multiple
scattering suppression is then achieved by an efficient cross
correlation algorithm working in real time with a temporal
resolution down to 0.2 seconds. Our approach allows access to the
extensive range of systems that show low-order scattering by
selective detection of the singly scattered light. Model
experiments on slowly relaxing suspensions of titanium dioxide in
glycerol were carried out to establish the validity range of our
approach. Successful application of the method is demonstrated up
to a scattering coefficient of more than $\mu_s = 5$~cm$^{-1}$ for the
sample size of $L=1$~cm.

\end{abstract}


\maketitle


\section{INTRODUCTION}

Dynamic Light Scattering (DLS) analyzes the intensity fluctuations
of light scattered from a medium in the weak scattering limit.
This is typically done by means of the normalized intensity
autocorrelation function (ACF):
\begin{equation}
g_2(q, \tau) = \frac{\langle I(q, t) I(q, t + \tau)
\rangle_t}{\langle I(q,
  t) \rangle_t^2},
\end{equation}
where $\langle \cdots \rangle_t$ denotes time averaging, $\tau$ is the lag time and $q$ is
the \emph{scattering wave number} or \emph{momentum transfer}
defined by scattering angle $\theta$, the wavelength in vacuum
$\lambda$ and the solvent refractive index $n$: $q = 4 \pi n
sin(\theta/2) / \lambda$. The measurable quantity $g_2(q, \tau)$
can be linked
 to actual physical microscopic properties by the normalized field autocorrelation
 function
$g_1(q, \tau)$ via the Siegert relation:
\begin{equation}
g_2(q, \tau) = 1 + \beta | g_1(q, \tau)|^2,
\label{eq:siegert}
\end{equation}
where the coefficient $\beta$ depends on the detection optics.

Quite generally the the field autocorrelation function $g_1(q,t)$ provides
access to thermally driven local dynamic properties on length
scales of the order $1/q$. A prominent example is the Brownian
motion of colloidal particles in a solvent such as water. For this
most simple case the normalized field correlation function can be
written as \cite{berne:dls}:
\begin{equation}
g_1(q, \tau) = exp(-D_0 q^2 \tau)=exp(-\tau/\tau_c), \label{eq:g1}
\end{equation}
where $\tau_c=1/{D q^2}$ is the relaxation time and $D_0$ is the
particle diffusion coefficient defined by the Stokes-Einstein
relation:
\begin{equation}
D_0 = \frac{k T}{6 \pi \eta R}, \label{eq:D0}
\end{equation}
with $\eta$ the solvent viscosity, $T$ is the sample temperature and $R$ the particle radius.
The equation \eqref{eq:g1} is widely used in dynamic light scattering
for the sizing of small particles.

An essential condition for traditional dynamic light scattering to
work is absence of multiple scattering. As soon as higher
order scattering becomes considerable (typically if transmission in
line-of-sight is lower than 95 per cent) the measured intensity
correlation function starts to deviate from the theoretical
expectations, leading to a faster decay. Furthermore, information
on the scattering wave number as well as on the scattering angle
$\theta$ is lost since the detected light is composed of
several scattering events with unknown momentum transfer.

The influence of multiple scattering can be reduced by decreasing
the concentration of the sample under study or the cell size or by
refractive-index matching \cite{megen:dynamic}. The latter one is
usually not possible without changing other sample properties.
Limitations to the size of the container are set by the optical
quality of the sample cells and by boundary effects. For
cylindrical cells minimal diameters of typically $1-3$ mm are used
whereas in flat or rectangular containers even smaller photon path
lengths are achieved \cite{urban:3ddls,lehner:cells}. To maximally
reduce the photons path lengths one can use fiber optical probes
\cite{thomas:fiber} directly immersed in the (liquid) sample. This
approach known as Fiber Optical Quasi Elastic Light Scattering
(FOQELS) has been applied in a number of recent studies (see, e.g.
Refs.~\cite{lilge:fiber,wiese:fiber,horn:fiber}). The application
of FOQELS is however limited to backscattering angles around
180$^\circ$  and moreover the interpretation of the data is often
complicated due to the incomplete suppression of multiple
scattering.

Quite a different way of actively dealing with multiple
scattering has been put forward over the last two decades. The
idea is to carry out two simultaneous DLS experiments with exactly
the same scattering vectors in the same scattering volume  and
analyze the time cross-correlation function. It has been clearly
shown that under proper conditions (see
Refs.\cite{phillies:supp,phillies:exp,schatzel:supp}) the
cross-correlation function equals the auto-correlation function
for single scattering within the range of experimental resolution.
Successful implementations of this scheme have been reported by
several groups. The techniques are called two color DLS (TCDLS)
\cite{schatzel:supp,pusey:suppression} and three dimensional DLS
(3DDLS) \cite{urban:3ddls,pusey:suppression} respectively and the
latter one is available commercially \cite{lsinstruments}.

\begin{figure}
\mbox{
      \subfigure[\label{fig:theorem}]{
    \includegraphics[width=0.9\linewidth]{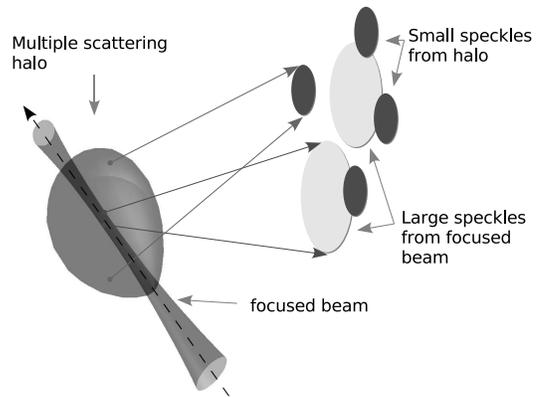}
} }
\mbox{
      \subfigure[\label{fig:scheme}]{
      \includegraphics[width=0.8\linewidth]{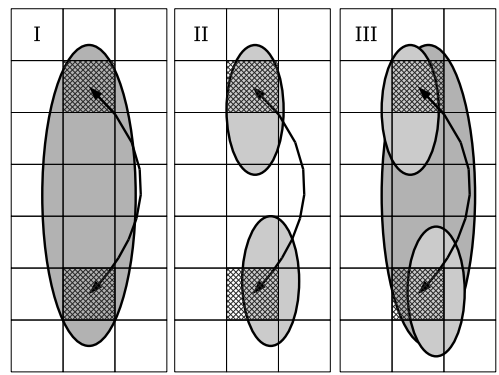}
}}
\caption[]{
    (a) Illustration of the van Cittert-Zernike theorem: small coherently illuminated areas (such as a focused laser beam)
  produces large correlated areas (speckles) and and vice versa: large coherence areas
  (halo from multiple scattering) produces small speckles. (b)
  Suggested suppression scheme using cross-correlation processing: I)
  intensity values measured within the same speckle are correlated II) two
  different speckles are uncorrelated III) for a superposition of small and large speckles intensities detected at a certain distance will
  be correlated only due to the larger speckles. \newpage}

\end{figure}

Another cross-correlation approach uses a single-beam two-detector
configuration ~\cite{meyer:supp,schroder:supp}. Suppression of multiple
scattering is based on a consequence of the van Cittert-Zernike
theorem \cite{goodman:so}, which states that intensity
correlations in an observation region are closely related to the
Fourier transform of the intensity distribution across the
source. This means that a small region of single scattering (e.g.
the volume of a focused beam) will produce large correlated areas
(speckles), whereas a comparably large halo of multiple scattered
photons will give rise to small speckles (see Fig.~\ref{fig:theorem}).  This is
reflected in the well known expression \cite{goodman:so} for the
speckle size $\delta x$ in the far-field geometry: $\delta x = \lambda z /
S$, where $z$ is the distance from the light emitting object to
the detector and $S$ is the lateral extension of the object along
one chosen direction. The consequences for a light scattering
cross-correlation experiment are obvious. Spatially resolved
detection of the scattered intensity will carry selective
information about the spatial distribution of light in the
scattering volume.

If we assume that the size of single-scattering volume $S_1$ is
equal to a average radius of beam cross-section of $S_1 \approx 20$~\textmu m
and the scattering mean free path be $l \approx  1$~mm then
the volume of double-scattering extends over roughly a \emph{25} times larger cross
section. In practice the dimension of the detected scattering
volume will determine $S_2$ both for weak and moderate multiple
scattering.

Scattered intensities measured in two points separated by distance
$\Delta x$
\begin{equation}
\frac{\lambda \cdot z}{S_1}  \ge \Delta x \gg \frac{\lambda \cdot
z}{S_2}, \label{eq:speckle_size}
\end{equation}
will thus be correlated only due to single scattering as shown in
the Fig.~\ref{fig:scheme} so the normalized intensity cross
correlation function (CCF) 
\begin{equation}
g^{\Delta x}_2(q, \tau) = \frac{\langle I(q, t, 0) I(q, t + \tau,
  \Delta x)
\rangle_t}{\langle I(q,  t, 0) \rangle_t \langle I(q,
  t, \Delta x) \rangle_t}
\end{equation}
will provide the proper estimate of auto correlation function of
singly scattered intensity. Such an approach has already been
successfully demonstrated by Meyer \emph{et al.} \cite{meyer:supp}
with the scheme based on cross-correlation of scattered intensities detected by two
spatially separated fibers.
While the underlying optical background (van Cittert - Zernike
theorem) is highly plausible it is more complicated to put forward
a detailed theoretical description since this requires modeling
of the low order scattering processes. Such treatment has been
derived by Lock \cite{lock:role} for the case of double scattering. He finds the
multiple scattering suppression ratio to be approximately
proportional to the speckle size ration $S_2/S_1$  if the
detectors are placed at the distance $\Delta x \approx \lambda \cdot
z / {S_1}$. The suppression ratio can be improved by choosing a larger
separation $\Delta x$ albeit at the cost of a decreased signal.
Choosing a large distance on the other hand might prove
unnecessary for small amounts of multiple scattering. It is due to
these practical difficulties, that the technically simpler
single-beam cross-correlation geometry is often considered
inferior to the two-beam realization (where a sample independent
accurate theoretical description is available).

Here we propose an extension of the single-beam cross-correlation
method that allows to overcome this shortcoming. Using a
charge coupled device (CCD) camera as a detector we can analyze speckle correlations and
adapt $\Delta x$ thus assuring single-scattering detection with
high accuracy. As we will show this flexibility together with the
intrinsically high statistical accuracy of multi-speckle detection
leads to a much improved performance of the single-beam
two-detector configuration while essentially preserving its
technical simplicity.

\section{EXPERIMENTAL SETUP}

\begin{figure}

  \begin{center}
    \begin{tabular}{c}
      \includegraphics[width=\linewidth]{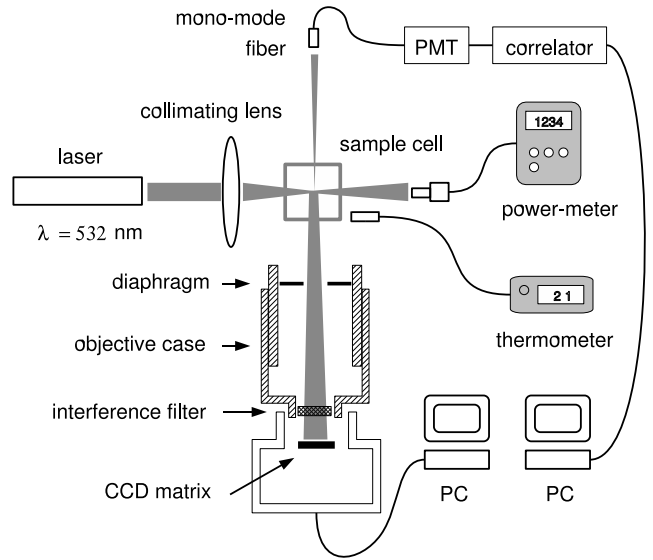}
    \end{tabular}
  \end{center}
  \caption[example]
      { \label{fig:setup}
        Experimental setup. Light scattered at an angle of
  90\textdegree~  inside the sample cell of thickness $L=1$ cm passes a
  diaphragm  and is collected by a CCD digital camera. In opposite direction light is collected by a mono-mode fiber and
 a photon counting module to be processed by a hardware correlator.
 The transmitted collimated intensity is measured to determine the scattering parameters $\mu_s$. }
\end{figure}

The fluctuations of scattered light intensity are monitored in a
traditional scattering geometry using a fixed scattering angle. As
a light source solid-state laser system (Verdi Coherent
Inc., USA) operating at the wavelength $\lambda = 532$~nm  is used. The
beam is strongly focused by a lens with focal length $f = 50$~mm
inside the sample cell which results in an average beam radius of
20~\textmu m inside the scattering volume. The focal spot was positioned in the center of a
rectangular quartz cell (Hellma GmbH, Germany) with inner base
dimensions of 10~\texttimes~10~mm and height 45~mm. The light
scattered from the sample is recorded simultaneously by
single-mode fiber connected to a photon counting module
(PerkinElmer, Canada) and analyzed with a digital correlator
(Correlator.com, USA) from one side and a CCD camera from the
opposite side. The CCD grayscale camera ``Pixelfly'' produced by PCO
Computer Optics GmbH, Germany is configured to operate in VGA mode
(640~\texttimes~480 pixels) with 12~bits resolution and 50~frames
per second speed with exposure time $\tau_0 = 20$~ms. This camera
is based on a Sony ICX414AL image sensor with the pixel size of
9.9~\texttimes~9.9~\textmu m. The dynamic range of analog/digital
conversion of the CCD signal is 68.7~dB (according to manufacturer
specifications). A diaphragm with an opening diameter of roughly
3.5~mm selects the range of scattering vectors seen by the CCD
matrix and also determines the effective scattering volume. The
distance for the beam center to the CCD matrix was $z = 11$ cm which
corresponds to the speckle size of ca. 14 \textmu m or 1.4 pixels.

Scattering angles covered by the CCD chip are found in the range
$\theta = 90^\circ\pm 3.15^{\circ}$. The corresponding scattering
vector is equal to $q = (24.59 \pm 0.67) \cdot 10^6$~m$^{-1}$ with
a deviation of the order of 2.75 \% from the average  90$^{\circ}$ angle.
An improved angular accuracy could be achieved by placing the
camera further away from the sample however at the cost of
decreased statistical accuracy due to the smaller number of
detected speckles. A better way to deal with this problem would be
the processing of an angle resolved digital image. However this
has not been realized in this study.

In parallel the intensity of the collimated beam transmitted
through the cell is measured by a laser power-meter FieldMax
(Coherent Inc., USA).  The temperature is monitored by a digital
thermometer with the probe placed close to the cell surface.

As a model system we studied the Brownian motion of  commercial  TiO$_2$
particles in pure (99.5\%) glycerol.  Solutions were sequentially
passed through filters with pore diameters 5~\textmu m and
1.2~\textmu m, respectively. Measurements on samples highly diluted
with water 
reveal a mean hydrodynamic diameter for TiO$_2$ particles of
293~$\pm$~15~\textmu m,
in qualitative agreement with electron micrographs.
For all measurements the temperature was in the range 20 $\pm$
0.5 $^\circ$C. Corresponding variations of solvent viscosity are
of the order of $\pm$ 5 \% \cite{handbook:64th}.  We select Glycerol as a
solvent to decrease the particle diffusion coefficient $D_0$ and
thus enable real-time detection and processing of the scattered
intensity fluctuations with a CCD camera \footnote{
We note that Glycerol is hygroscopic and it is thus difficult to
know precisely the exact water content. As a consequence a small
uncertainty remains with respect to the solvent viscosity. We have
assured, however, that all samples were prepared under identical
conditions. Thus any systematic shift in the solvent viscosity
will not affect the results of our study.}.

Typically $10^4-10^5$ frames were collected corresponding to a
measurement time of  $15-30$ minutes. The actual number of
collected correlation coefficients depends on the processing
scheme (see the next section for details) but usually is of the
order of $10^8$ for the correlation coefficient of smallest delay
time.

The turbidity of the sample was characterized by means of the
scattering coefficient $\mu_s $. It is proportional to the
particle density $\rho$ and the scattering cross section
$\sigma_s$: $\mu_s= \rho \sigma_s$. Due to the unknown amount of
particles lost in the filtering process however no accurate
density values are available. Since even fairly small particle
densities lead to considerable multiple scattering we can
nevertheless safely assume to work in the highly dilute limit.
Experimentally  $\mu_s$  can be estimated from the transmitted
intensity $I_t$ according to Lambert-Beer's law for the case of
non-absorbing particles: $I_t / I_0 = A exp(-\mu_s L)$, where
$I_0$ is the intensity incident on the cell, $L$ is the cell
thickness, and $A$ describes loss and deflection of intensity at the surface
of the cell. The latter is independent of the particle density and
was estimated from the transmission coefficient of a cell
containing pure glycerol.


\section{PROCESSING TECHNIQUES}

A key element of our study is the optimization of multi-speckle
detection and processing schemes. Our goal is to combine
intelligent optical realizations with the power of modern parallel
processing of a large amount of data acquired by a digital camera.
Such optimized data analysis can be both achieved in the time and
and the space domain. As the first step we will discuss the application
of the multi-tau scheme as developed by Sch\"atzel to our
multi-speckle analysis (previous realization for digital cameras
are described e.g. \cite{cipelletti:ultralow}). Secondly, we will
show that a conceptually very similar approach can be used in the
space domain in order to improve both the multiple scattering
suppression efficiency and the processing speed of our CCD
detection scheme.
\subsection{Multi-tau correlation scheme}
\begin{figure}
\mbox{
      \subfigure[\label{fig:mtau}]{
      \includegraphics[width=0.6\linewidth]{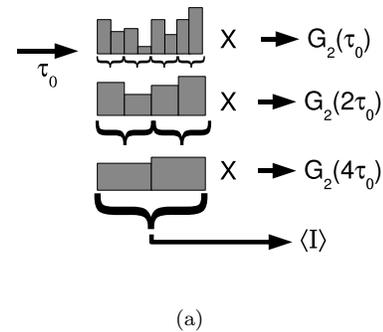}

}} \quad
\mbox{
      \subfigure[\label{fig:mtau_scheme}]{
      \includegraphics[width=0.95\linewidth]{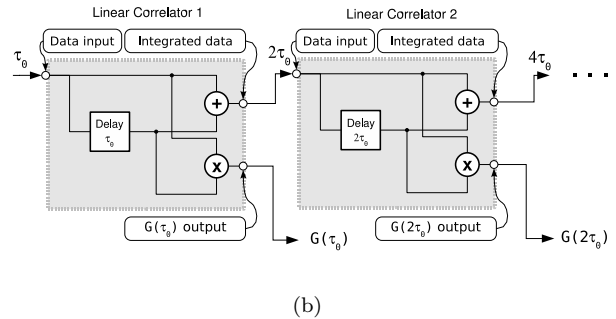}
}
      }
\caption{(a) Principles of the multi-tau correlation scheme.
Sequential intensity values $I(t)$ are integrated for the
      computation of larger lag times.(b) Simplified object scheme to illustrate
      the cascading of linear correlators used for the realization
        of the multi-tau correlation scheme.}
\end{figure}

The multi-tau correlation scheme was originally proposed by
Sch\"{a}tzel \cite{schatzel:photon} to increase the accuracy of
hardware correlators for large lag times. In most cases the linear
spacing of lag times is not required  to analyze single scattering
correlation functions. Instead one can increase the distance
between lag points in the correlation function for large lag
times, thus saving valuable processing time. Such an approach can
be easily and efficiently realized with a sequential doubling of
the effective exposure time, which furthermore improves
statistical accuracy.

 Let us suppose that the camera provides
data with an initial exposure time equal to $\tau_0$ with $1 /
\tau_0$ Hz frequency which implies the absence of delays between
the collection of sequential images. Then the doubling of the
exposure time to $2 \tau_0$ can be realized by integrating two
sequential values of data obtained with exposure time $\tau_0$. In
the same way the effective exposure time can be increased to $4
\tau_0$ and so on as it is shown in Fig.~\ref{fig:mtau}. From the
obtained time series of intensity fluctuations with different
exposure times the correlation coefficients can be calculated with
a simple linear scale processing.

In terms of software realization this is an ideal case for an
object-oriented approach. The main object in this scheme is an
Elementary Linear Correlator shown in Fig.~\ref{fig:mtau_scheme},
which receives the sequential data point $I(i)$ as an input at
each cycle $i$, multiplies the new value with previous data points
$I(i - k)$ for different linearly spaced integer delays $k =
0,1,2,...$ and updates the mean values of the corresponding
products $\langle I(i) I(i - k) \rangle$. Every second cycle the
correlator  estimates the mean integrated value of intensity for
two sequential time steps $[I(2n) + I(2n+1)] / 2$ (where $n =
0,1,2,...$) and sends it to the next linear correlator, thus 
forming a cascading line of correlators. Since the data rate on a
sequential correlator is half of the data rate compared to the
preceding one, the evaluation period doubles at every correlator.
 Extra effort has to be taken to optimize the cycles with respect to limitations in computational power.

The elementary correlator unit used in this study was designed to
calculate the cross-correlation products together with the
auto-correlation. ACF coefficients are calculated on every pixel
individually to obtain local values of $\langle I(x,t) I(x,
t+\tau) \rangle_t$ and $\langle I(x, t) \rangle_t$. For the CCF
with separation $\Delta x$ we obtain $\langle I(x, t) I(x +\Delta
x, t+\tau) \rangle_t$ and $\langle I(x, t) \rangle_t$ and $\langle
I(x+\Delta x, t) \rangle_t$. These values are accumulated in each
correlator for further spatial averaging.

The scheme described here allows us to evaluate the data in real
time and therefore does not set any limitations on the duration of the
measurements.

\subsection{Binning technique}
The low dynamic range of CCD cameras in comparison to photodiodes
or photo-multipliers used in previous studies
\cite{meyer:supp,schatzel:supp} is the main challenge for their
utilization in a cross-correlation scheme which requires resolving
the small signal from single scattering hidden by a dominating
signal from multiple scattering. The problem can be partially
solved with an original binning technique developed in the course
of this study. If the size of a speckle from single scattering
exceeds the area of several pixels it can be approximated with
integral values of these pixels intensities. Let us call the bin
or meta-pixel the area $S_x \times S_y$ pixels represented by a
single intensity value obtained by integration (or floating-point
averaging) of intensities of included pixels. Due to multiple
sampling the noise of measurements will be reduced 
by factor $\sqrt{S_x S_y}$ \cite{dspguide} and thus the dynamic range defined
as $DR = 20 \log_{10} SNR$~dB, where $SNR$ is a signal-to-noise
ratio, will increase on $10 \log_{10} S_x S_y$~dB. For example
for a $10 \times 2$ window the dynamic range of our camera model
is increased up to 81.7~dB or a $SNR = 1.22 * 10^4$. The loss of
spatial resolution reduces the statistical accuracy and intercept
$\beta$ only if the binning area $\sqrt{S_x \times S_y}$ is
comparable or larger than the coherence area $\sqrt{S_1 \times
S_2}$. As a consequence binning leads to partial suppression of
multiple scattering by averaging out small speckles. Binning
introduces a
 two-dimensional filtering with a box-like kernel function which
 will reduce more efficiently the high-frequency spatial fluctuations
 (multiple scattering speckles) than the lower-frequency
 fluctuations that are connected to single-scattering speckles.
In other words the binning technique in our cross-correlation
approach can be considered a spatial analogue of the multi-tau
technique in the time domain described above.

\subsection{Multi-speckle averaging technique}

The ability of cameras to register a large number of independently
fluctuating speckles simultaneously
can be very efficiently used to increase the statistical accuracy of a
measurement by
averaging the data along the bins. For the case of independent speckles
different pixels or bins can be treated as separate photodetectors and
thus their measurements can be processed all together to estimate the
correlation function. Actually two possible ways of dealing with this
data exist. As it was mentioned above correlators collect the
time-averaged products $\langle I(x,t) I(x, t+\tau) \rangle_t$ as well
as the mean intensity $\langle I(x,t) \rangle_t$ for certain bins. Thus
the further averaging followed by a normalization with the mean
intensity can be performed with products and mean intensities as
was proposed in the original multi-speckle scheme \cite{kirsch:multispeckle,knaebel:aging}:

\begin{equation}
g^{ad}_2(\tau) = \frac{\langle \langle I(x,t) I(x, t+\tau) \rangle_t
  \rangle_x}{\langle \langle I(x,t) \rangle_t \rangle_x^2} = \frac{\langle \langle I(x,t) I(x, t+\tau) \rangle_t
  \rangle_x}{\overline{I}^2},
\end{equation}
where $\langle \ldots \rangle_x$ denotes the averaging along the entire
two-dimensional CCD matrix and $\overline{I}^2 = \langle \langle I(x,t) I(x, t+\tau) \rangle_t
  \rangle_x = \langle \langle I(x,t) I(x, t+\tau) \rangle_x  \rangle_t$ is a mean value of intensity averaged over time and space. This
``average-and-divide'' sequence can be applied for the ideal case
of a uniform illumination of the CCD matrix, i.e. when the mean
intensity $\langle I(x,t) \rangle_t$ is not a function of $x$.
Because for $\tau \rightarrow \infty$, one finds $I(x,t), I(x,
t+\tau) \rightarrow \langle I(x,t) \rangle_t$ and $\langle I(x,t)
I(x, t+\tau) \rangle_t \rightarrow \langle I(x,t) \rangle_t^2$, so
for this processing scheme

\begin{equation}
\lim_{\tau \rightarrow \infty} g_2(\tau) = \frac{\langle [
  I(x,t) \rangle_t^2 \rangle_x}{\overline{I}^2} =
\frac{\langle [ \langle I(x,t) \rangle_t - \overline{I}]^2 \rangle_x}{\overline{I}^2} + 1 =
   K + 1,
\end{equation}
where $K = \langle [ I(x,t) - \overline{I}]^2 \rangle_x /
 \overline{I}^2$ is the contrast of the picture that would be obtained
 with infinite exposure time. For the case of uniform illumination
$K = 0$ due to the fact that $\langle I(x,t) \rangle_t = \overline{I}$
 all along the detector matrix and thus $g_2(\tau \rightarrow \infty)$  will approach 1 so the
 normalized  field autocorrelation function $g^2_1(\tau) = g_2(\tau) - 1$ will
 decay to zero. But even for a slightly non-uniform illumination when
 $K > 0$ the non-zero additive component appears in the measured values
 $g^2_1(\tau)$.

To overcome this problem the original multi-speckle scheme was
somewhat modified in our study by  normalization of the locally
estimated correlation coefficients, and averaging only after that
(``divide-and-average'' sequence):
\begin{equation}
g^{da}_2(\tau) = \left \langle \frac{\langle I(x,t) I(x, t+\tau) \rangle_t}
{\langle I(x,t) \rangle_t ^2} \right \rangle_x.
\end{equation}
Still this requires the deviation of the mean intensity along the
matrix to be small in comparison to the noise level. For a
perfectly uniform illumination of the matrix both algorithms
provide the same result.

\section{RESULTS AND DISCUSSION}

We have carried out a series of experiments using samples of
different scattering strength, hence different amounts of multiple
scattering, ranging from the dilute limit $\mu_s L=0.1$ to a
regime $\mu_s L=5.92$ where the transmitted beam is attenuated to
only $0.27\%$ of its initial intensity.
\begin{figure}
  \begin{center}
    \begin{tabular}{c}
      \includegraphics[width=\linewidth]{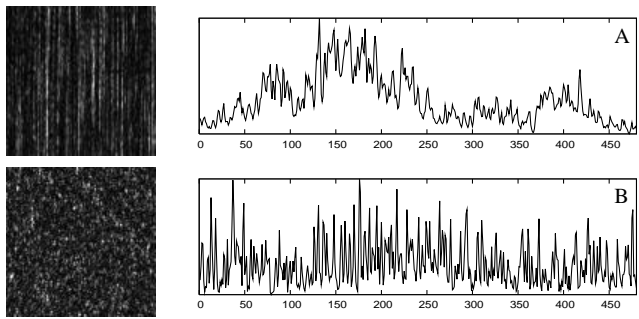}
    \end{tabular}
  \end{center}
  \caption { \label{fig:speckle_traces}
   Left: Representative area taken from the recorded images for a weakly scattering  (A) and a strongly scattering (B) sample. Right:
   Spatial intensity fluctuations along the vertical columns of the CCD matrix. Sample A: $\mu_s$ = 0.101
    cm$^{-1}$, estimated speckle size $\delta x$ = 28.02 pixels, sample B:  $\mu_s$ = 5.92
    cm$^{-1}$, $\delta x$ = 0.76 pixels}
\end{figure}

\subsection{Spatial intensity correlations in the speckle pattern}

As soon as multiple scattering effects become considerable the detected
intensity distribution will consist of a distribution of
correlation lengths. The longest correlation length, $\delta
x=\lambda z \approx 300$ \textmu m $\approx 30$ pixels can be associated
with single scattering whereas higher order scattering leads to
smaller speckles. Figure~\ref{fig:speckle_traces} illustrates this
by showing the intensity fluctuations along the CCD matrix
columns. For the case of single scattering sample
($\mu_s$~=~0.1~cm$^{-1}$, upper plot) the anisotropy of the
speckle pattern can be clearly seen. The reason for the
anisotropy is the horizontal confinement of the incident focused beam that
defines the scattering volume in this regime. The scale of
intensity fluctuations along the ``$x$'' dimension are of the
order of tens of pixels so the correlated areas of intensity or
speckles are large. With increasing scattering coefficient $\mu_s$
the fluctuations become more pronounced and the correlation length
decreases. The horizontal speckle-size is always given by the
dimensions of the detected scattering volume both in the single
and multiple scattering case.

\begin{figure}
  \begin{center}
    \begin{tabular}{c}
      \includegraphics[width=\linewidth]{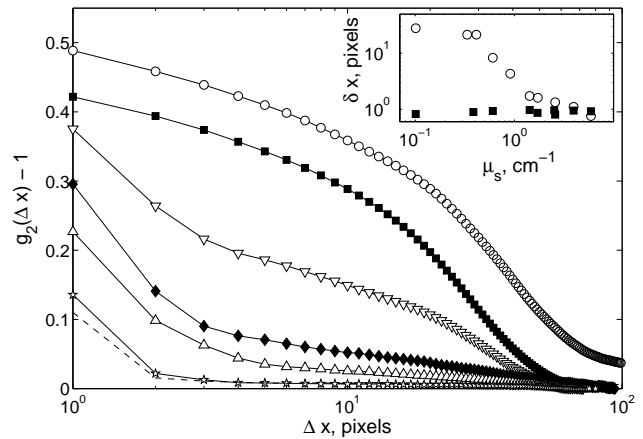}
    \end{tabular}
  \end{center}
  \caption
      { \label{fig:spatial_acfs}
    Spatial auto-correlation for different sample turbidities
    as a function of vertical pixel separation $\Delta x$:
    ($\circ$) $\mu_s$ = 0.10 cm$^{-1}$,
    ($\blacksquare$) $\mu_s$ = 0.39 cm$^{-1}$,
    ($\bigtriangledown$) $\mu_s$ = 0.88 cm$^{-1}$,
    ($\blacklozenge$) $\mu_s$ = 1.43 cm$^{-1}$,
    ($\bigtriangleup$) $\mu_s$ = 2.54 cm$^{-1}$,
    ($\star$) $\mu_s$ = 5.92 cm$^{-1}$.
    Dashed line ($- - -$) represents the same estimated in horizontal dimension.
     Inset shows speckle sizes $\delta x$ ($\circ$) and $\delta y$ ($\blacksquare$) as a function of $\mu_s$
    for both vertical and horizontal
    separations respectively. The maximum amplitude of the
    correlation function is limited to about 0.5 since the
    speckle size in $y$-direction is comparable to the pixel size}
\end{figure}

For a quantitative estimation of the actual speckle size the
normalized spatial auto-correlation function can be used:
\begin{equation}
g^s_2(\Delta x) = \frac{\langle I(x) I(x + \Delta x)\rangle_{x,t}}{\langle
  I(x) \rangle_{x,t} \langle
  I(x + \Delta x) \rangle_{x,t}},
\end{equation}
where $\langle \cdots \rangle_{x,t}$ denotes the time and space
averaging. In practice the averaging is performed through all pixel
pairs separated vertically by a distance $\Delta x$. A total of 8
independent speckle images is analyzed with an exposure time of
20~ms. Temporal fluctuations are much slower, typically of the
order of 0.3 seconds or more, and could thus be neglected. Estimated
values for the mean speckle size $\delta x, \delta y$ can 
be defined by the condition $g^s_2(\Delta x,\Delta y \ll \delta y)
- 1=1/e$, $g^s_2(\Delta x \ll \delta x,\Delta y) - 1=1/e$. Since
speckles in vertical direction are always small, the limit $\Delta
y \ll \delta y$ cannot be reached in our setup and we have thus
 calculated the correlation length in $\Delta x$
direction using the correlation function normalized with respect to $g_2^s (\Delta x = 1, \Delta y = 1)$. Spatial ACFs
for samples of various turbidities are shown in
Fig.~\ref{fig:spatial_acfs} together with the estimated speckle
sizes. In vertical direction, with the increase of $\mu_s$, the
speckle size decreases rapidly. In the multiple scattering limit
the speckles are almost isotropic with a size of approximately $1
\times 1$ pixels which means that the initially focused beam is dispersed into
a diffuse scattering cloud.

The spatial resolution in our experiments allows direct 
estimation the relative weight of single scattering contributions.
For separations much larger than 3 pixels the cross correlation
signal will be dominated by single scattering. A separation
comparable to the size of a single scattering speckle, $\Delta x
\approx 30$ pixels is a good compromise value separating single
from multiple scattering.

Note that using a CCD camera we are however not restricted to fixed detector
positions as in previous studies. Smaller separations $\Delta x$
could be used in the weak scattering limit whereas a gradual
increase of $\Delta x $ in the strong scattering limit
simultaneously optimizes multiple scattering suppression and the
signal-to-noise ratio.

\begin{figure}
    \begin{tabular}{c}
      \includegraphics[width=\linewidth]{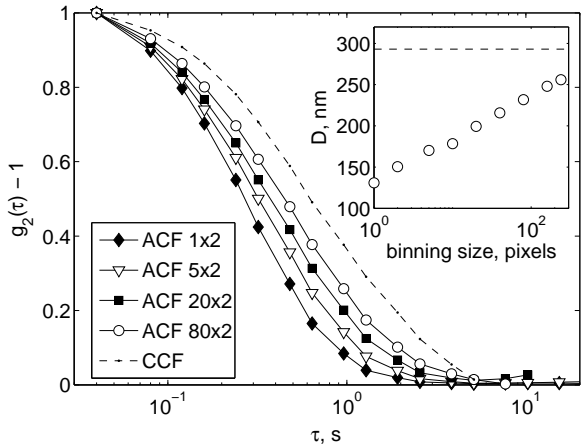}
    \end{tabular}
  \caption[example]
      { \label{fig:binning_effect}
        Auto correlation functions obtained with different binning areas
        together with the cross-correlation function. As the
        vertical size of the binning area increases the ACF approaches the CCF indicating partial but
        not sufficient suppression of multiple scattering . The inset shows the estimated particle size ($\circ$) as a function of
        binning size. The actual size is indicated with by a dashed line. }
\end{figure}

\subsection{Pixel binning}

Instead of pixel by pixel cross correlation we can apply the
superior binning approach described before. Proper selection of the
binning areas can increase the signal-to-noise ratio and also
partially filter out the small multiple scattering speckle. The
latter effect is demonstrated in Fig.~\ref{fig:binning_effect} for
different numbers of binning pixels in the vertical direction.
Since the horizontal speckle size is constant the binning in this
dimension was selected to be 2
pixels.

Auto correlation functions calculated from such larger areas are
found increasingly closer to the single scattering function as the
binning area is increased. However complete suppression is
difficult to achieve since multiple scattering suppression scales
linearly with the number of pixels whereas cross correlation
scales exponentially with pixel separation.  Another disadvantage
of very large binning areas is the reduction of the statistical
ensemble. Ideally the binning area is chosen somewhat smaller than
the size of single scattering speckle in order to retain a high
number of independent speckles. In our case the best choice in
$x$-direction is in the range of typically 5-20 pixels as it can be
seen from Fig.~\ref{fig:spatial_acfs}.

\begin{figure}
  \begin{center}
    \begin{tabular}{c}
      \includegraphics[width=\linewidth]{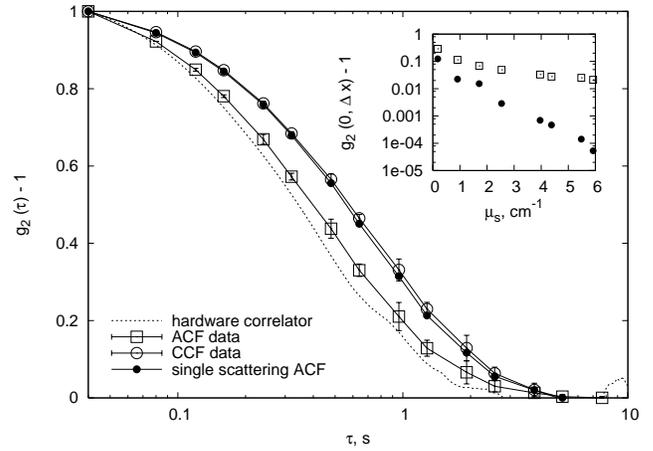}
    \end{tabular}
  \end{center}
  \caption
      { \label{fig:ccf_complete}
        Correlation functions obtained by different means for
        an essentially  multiple-scattering sample ($\mu_s = 3.93$ cm$^{-1}$).
	The ACF obtained from the hardware correlator
	($- - -$), the ACF from CCD detection with no binning
        ($\square$) and the
        CCF with a separation $\Delta x = 40$ pixels ($\circ$)  are
	shown together with the  ACF
        obtained for singly scattering sample ($\bullet$). Inset shows
        the intercepts for ACF ($\square$) and CCF  with separation
        $\Delta x = 40$ ($\bullet$)
        pixels as functions of $\mu_s$.}  
\end{figure}

\subsection{Intensity Correlation Functions}

Fig.~\ref{fig:ccf_complete} shows the correlation function for
one of the most strongly scattering samples using different
measurement schemes. With our optimized cross-correlation ($\Delta
x = 40$ pixels) and binning approach ($20 \times 2$) the measured
correlation function perfectly agrees with the single scattering
result.

\begin{figure}
  \begin{center}
    \begin{tabular}{c}
      \includegraphics[width=\linewidth]{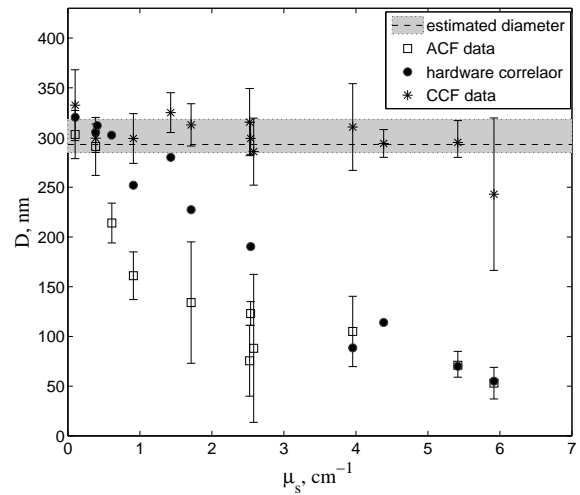}
    \end{tabular}
  \end{center}
  \caption{ \label{fig:accuracy}
        Particle size $D$ determined by different methods as a
            function of sample turbidity. The CCF diameter remains unchanged up to the limit strong multiple scattering. T.
            ACF from the hardware correlator ($\bullet$) and from ACF without binning
        ($\square$) show a rapid decrease as the
            scattering coefficient $\mu_s$ increases.The shaded
            area indicates the experimental value expected for single
        scattering (including the uncertainty in particle size and
        solvent viscosity)}
\end{figure}

A set of measurements has been carried out and the measured
particles diameters are presented in Fig.~\ref{fig:accuracy} as a
function of scattering coefficient $\mu_s$. For the case of dilute
samples ($\mu < 0.5$ cm$^{-1}$) all processing schemes yield the
same results within the experimental error. Increasing $\mu_s$
results in smaller apparent values of the particle size for the
case of autocorrelation measurements (ACF). At the same time the
particle diameter obtained using our multiple scattering
suppression scheme (CCF processing) stays the same for all $\mu_s
< 5.5$ cm$^{-1}$ corresponding to $\mu L = 5.5$. Only for the most
turbid sample $\mu_s \approx 5.92$~cm$^{-1}$  a noticeable
difference is observed due to the loss of the single scattering
signal and the unavoidable increase of error.


Finally we would like to comment on the performance of the
multi-speckle averaging scheme implemented in our approach.
Neglecting any angular dependence we are treating all $x$-columns
equally. As we have seen the size of a single scattering speckle
roughly amounts to $30 \times 2$ pixels. Since the full CCD chip
has $640 \times 480$ pixels we detect approximately 5000
independent speckles. This number can be increased only if the
diameter of the incident beam is increased, but at the expense
of a inferior performance in multiple scattering suppression.

\section{CONCLUSION AND OUTLOOK}

CCD-camera based light scattering can be used to efficiently
suppress the undesirable effects of multiple scattering. Our
single-beam cross correlation scheme combines for the first time
the advantages of multiple-scattering suppression and
multi-speckle detection. The latter reduces the collection time
for slowly evolving samples dramatically and provides the
necessary statistical accuracy to detect even small cross
correlation signals deep in the multiple scattering regime.
Successful application of these combined processing schemes has
been demonstrated for sample turbidities as large as $\mu_s L
\approx 5$. We think that our relatively simple approach, heavily
relying on a flexible processing scheme, can be very useful for a
variety of applications for the study of complex fluids and soft
materials. With the maximum frame rate available in our
experiments the most appropriate areas of application can be found
in the field of slowly relaxing systems such as glasses and gels
or dense surfactant solutions, to name a few. But at the current
rate of development of image sensors and computer hardware much
smaller lag times seem to be feasible in a near future. Already
now, with the use of modern fast  complementary metal-oxide semiconductor
cameras with embedded
programmable digital signal processors the real-time on-camera calculation of the
cross-correlation functions is theoretically possible which in
turn would tremendously improve the time resolution.

\begin{acknowledgments}
Part of this work was supported by KTI/CTI program TopNano21 and Nestl\'{e} Research Center, Lausanne.
Support of the Swiss National Science Foundation is greatly acknowledged.
\end{acknowledgments}


\newpage

\section*{List of Figure Captions}

\noindent Fig. 1. (a) Illustration of the van Cittert-Zernike theorem: small coherently illuminated areas (such as a focused laser beam)
  produces large correlated areas (speckles) and and vice versa: large coherence areas
  (halo from multiple scattering) produces small speckles. (b)
  Suggested suppression scheme using cross-correlation processing: I)
  intensity values measured within the same speckle are correlated II) two
  different speckles are uncorrelated III) for a superposition of small and large speckles intensities detected at a certain distance will
  be correlated only due to the larger speckles.

\noindent Fig. 2.         Experimental setup. Light scattered at an angle of
  90\textdegree~  inside the sample cell of thickness $L=1$ cm passes a
  diaphragm  and is collected by a CCD digital camera. In opposite direction light is collected by a mono-mode fiber and
 a photon counting module to be processed by a hardware correlator.
 The transmitted collimated intensity is measured to determine the scattering parameters $\mu_s$. 

\noindent Fig. 3.
(a) Principles of the multi-tau correlation scheme.
Sequential intensity values $I(t)$ are integrated for the
      computation of larger lag times.(b) Simplified object scheme to illustrate
      the cascading of linear correlators used for the realization
        of the multi-tau correlation scheme.

\noindent Fig. 4.
   Left: Representative area taken from the recorded images for a weakly scattering  (A) and a strongly scattering (B) sample. Right:
   Spatial intensity fluctuations along the vertical columns of the CCD matrix. Sample A: $\mu_s$ = 0.101
    cm$^{-1}$, estimated speckle size $\delta x$ = 28.02 pixels, sample B:  $\mu_s$ = 5.92
    cm$^{-1}$, $\delta x$ = 0.76 pixels

\noindent Fig. 5.
    Spatial auto-correlation for different sample turbidities
    as a function of vertical pixel separation $\Delta x$:
    ($\circ$) $\mu_s$ = 0.10 cm$^{-1}$,
    ($\blacksquare$) $\mu_s$ = 0.39 cm$^{-1}$,
    ($\bigtriangledown$) $\mu_s$ = 0.88 cm$^{-1}$,
    ($\blacklozenge$) $\mu_s$ = 1.43 cm$^{-1}$,
    ($\bigtriangleup$) $\mu_s$ = 2.54 cm$^{-1}$,
    ($\star$) $\mu_s$ = 5.92 cm$^{-1}$.
    Dashed line ($- - -$) represents the same estimated in horizontal dimension.
     Inset shows speckle sizes $\delta x$ ($\circ$) and $\delta y$ ($\blacksquare$) as a function of $\mu_s$
    for both vertical and horizontal
    separations respectively. The maximum amplitude of the
    correlation function is limited to about 0.5 since the
    speckle size in $y$-direction is comparable to the pixel size

\noindent Fig. 6.
        Auto correlation functions obtained with different binning areas
        together with the cross-correlation function. As the
        vertical size of the binning area increases the ACF approaches the CCF indicating partial but
        not sufficient suppression of multiple scattering . The inset shows the estimated particle size ($\circ$) as a function of
        binning size. The actual size is indicated with by a dashed line. 

\noindent Fig. 7.
        Correlation functions obtained by different means for
        an essentially  multiple-scattering sample ($\mu_s = 3.93$ cm$^{-1}$).
	The ACF obtained from the hardware correlator
	($- - -$), the ACF from CCD detection with no binning
        ($\square$) and the
        CCF with a separation $\Delta x = 40$ pixels ($\circ$)  are
	shown together with the  ACF
        obtained for singly scattering sample ($\bullet$). Inset shows
        the intercepts for ACF ($\square$) and CCF  with separation
        $\Delta x = 40$ ($\bullet$)

\noindent Fig. 8
        Particle size $D$ determined by different methods as a
            function of sample turbidity. The CCF diameter remains unchanged up to the limit strong multiple scattering. T.
            ACF from the hardware correlator ($\bullet$) and from ACF without binning
        ($\square$) show a rapid decrease as the
            scattering coefficient $\mu_s$ increases.The shaded
            area indicates the experimental value expected for single
        scattering (including the uncertainty in particle size and
        solvent viscosity)

\newpage

\end{document}